\documentclass[12pt,preprint]{aastex}
\def\ms{m~s$^{-1}$}
\def\ks{km~s$^{-1}$}
\def\msini{$M_P\sin{i}$}

\def\vsini{$V_{\rm rot}\sin{i}$}
\def\msun{$M_{\odot}$}
\def\mjup{$M_{\rm Jup}$}
\def\rsun{$R_{\odot}$}
\def\lsun{$L_{\odot}$}
\def\chisq{$\sqrt{\chi^2_\nu}$}

\def\feh{[Fe/H]}
\def\rphk{$\log R^\prime_{HK}$}

\def\npllet{two}

\def\pC{1208}
\def\peC{30}
\def\pyearsC{3.309}

\def\tpC{2453102}
\def\tpeC{100}
\def\eC{0.146}
\def\eeC{0.08}
\def\omC{204}
\def\omeC{30}
\def\kC{24.0}
\def\keC{1}

\def\msiniC{1.8}
\def\arelC{2.7}
\def\rmsC{6.0}
\def\chisC{0.97}
\def\nobsC{48}
\def\mstarC{1.80}
\def\bvC{0.996}
\def\vmagC{4.79}
\def\mvC{2.32}
\def\vsiniC{1.5}
\def\ageC{2.5}
\def\rstarC{4.71}

\def\lstarC{12.3}

\def\teffC{4970}
\def\loggC{3.47}
\def\feC{+0.15}

\def\dC{31.1}

\def\pA{412.6}
\def\peA{4}

\def\tpA{2453330}
\def\tpeA{130}
\def\eA{0.027}
\def\eeA{0.04}
\def\omA{29}
\def\omeA{50}
\def\kA{32.7}
\def\keA{2}

\def\msiniA{1.7}
\def\arelA{1.3}
\def\rmsA{6.6}
\def\chisA{1.01}
\def\nobsA{31}
\def\mstarA{1.64}
\def\bvA{0.943}
\def\vmagA{5.97}
\def\mvA{2.48}
\def\vsiniA{1.5}
\def\ageA{2.2}
\def\rstarA{4.30}
\def\rstareA{0.070}
\def\lstarA{10.5}
\def\lstareA{0.050}
\def\teffA{5010}
\def\loggA{3.47}
\def\feA{+0.050}

\def\dA{50.0}

\def\rhkC{-5.40}
\def\shkC{0.10}
\def\rhkA{-5.34}
\def\shkA{0.11}

\def\starA{HD\,167042}
\def\teffA{5020}
\def\teffeA{75}
\def\loggA{3.52}
\def\loggeA{0.08}
\def\feA{+0.05}
\def\feeA{0.06}
\def\vsiniA{2.5}

\def\starC{$\kappa$\,CrB}
\def\teffC{4960}
\def\teffeC{70}
\def\loggC{3.45}
\def\loggeC{0.09}
\def\feC{+0.14}
\def\feeC{0.05}
\def\vsiniC{3.0}

\begin{document}
\title{Retired A Stars and Their Companions II: Jovian planets
  orbiting $\kappa$~Coronae~Borealis and \starA~\altaffilmark{1}}  

\author{ John Asher Johnson\altaffilmark{2,3}, 
  Geoffrey W. Marcy\altaffilmark{3},
  Debra	A. Fischer\altaffilmark{4}, 
  Jason T. Wright\altaffilmark{3},
  Sabine Reffert\altaffilmark{5},
  Julia M. Kregenow\altaffilmark{3},
  Peter K. G. Williams\altaffilmark{3},
  Kathryn M. G. Peek\altaffilmark{3}
}

\email{johnjohn@ifa.hawaii.edu}

\altaffiltext{1}{Based on observations obtained at the Lick
  Observatory, which is operated by the University of California.}
\altaffiltext{2}{Department of Astronomy, University of California,
Mail Code 3411, Berkeley, CA 94720}
\altaffiltext{3}{Institute for Astronomy, University of Hawaii,
  Honolulu, HI 96822}
\altaffiltext{4}{Department of Physics \& Astronomy, San Francisco
  State University, San Francisco, CA 94132}
\altaffiltext{5}{ZAH-Landessternwarte, K\"onigstuhl 12, 69117
  Heidelberg, Germany}

\begin{abstract}
We report precise Doppler measurements of \npllet\ stars, obtained at
Lick Observatory as part of our search for planets orbiting
intermediate--mass subgiants. Periodic variations in the radial
velocities of both stars reveal the presence of substellar
orbital companions. These two stars are notably massive with stellar
masses of \mstarC~\msun\ and 
\mstarA~\msun, respectively, indicating that they
are former A--type dwarfs that have evolved off of the main sequence
and are now K--type subgiants. The planet
orbiting \starC\ has a minimum 
mass \msini~$=$~\msiniC~\mjup, eccentricity $e = \eC$ and a \pC~day
period, corresponding to a semimajor axis $a = \arelC$~AU. The planet
around \starA\ has a minimum
mass \msini~$=$~\msiniA~\mjup\ and a \pA~day orbit, corresponding to a
semimajor axis $a = \arelA$~AU. The eccentricity of \starA\,b
is consistent with circular ($e = \eA \pm \eeA$), adding to the
rare class of known exoplanets in long--period, circular
orbits similar to the Solar System gas giants. Like all of the
planets previously discovered around evolved A stars, \starC b and
\starA b orbit beyond 0.8~AU. 
\end{abstract}


\section{Introduction}

Most of what is known about planets outside of our
Solar System comes from Doppler surveys of Sun--like stars, with
spectral types ranging from K0V to F8V and 
masses between 0.8 and 1.2~\msun\ \citep{valenti05, butler06,
  takeda06}. However, recent results from planet searches 
around low--mass M dwarfs \citep[][]{bonfils05b,endl06,butler06b} and evolved,
intermediate--mass stars \citep{reffert06, johnson06b, sato07, nied07,
  lovis07} have 
expanded into the frontiers at either end of the stellar mass
range. These surveys have begun to reveal important
relationships between stellar mass and the properties of exoplanets. For
example, planets around ``retired'' (evolved) A--type stars reside
preferentially in wide orbits, with semimajor axes $a \gtrsim 0.8$~AU
\citep{johnson07} 
and appear to have larger minimum masses than planets around Sun--like
stars \citep{lovis07}. Also, the occurrence rate of Jovian planets
(\msini$\geq 0.5$~\mjup, $a \leq 2.5$~AU)
increases with stellar mass, rising from $< 2$\% around M
dwarfs, to approximately 9\% around F and A stars \citep{johnson07b}.

The effects of stellar mass on other characteristics of
exoplanets, such as orbital eccentricity and 
multiplicity, will become evident as the sample of planets 
around intermediate--mass stars grows. Here, we report the
detection of two Jovian planets orbiting stars with masses of
\mstarA~\msun\ and \mstarC~\msun. These planet detections come from our
sample of intermediate--mass subgiants that we have been monitoring at
Lick and Keck Observatories \citep{johnson06b}. We describe our
observations in \S~\ref{observations}. In \S~\ref{stellar} we present
our data, and describe 
the characteristics of the host stars and the orbits of their
planets. We conclude in \S~\ref{summary} with a summary of our results
and a discussion of the properties of planets orbiting evolved A--type
stars. 

\section{Observations and Doppler Measurements}
\label{observations}

We have been monitoring a sample of 159 evolved stars at Lick
and Keck Observatories for the past 3.5 years. The stars were selected
from the \emph{Hipparcos} catalog based on the color and
magnitude criteria described by \citet{johnson06b}, namely 
$0.5 < M_V < 3.5$, $0.55 < B-V < 1.0$, and $V \lesssim 7.6$. We
exclude from our sample clump giants with $B-V > 0.8$ and $M_V < 2.0$,
as well as stars within 1~mag of the mean \emph{Hipparcos} main
sequence, as defined by \citep{wright05}. These criteria allow us to
take advantage of the widely spaced, nearly parallel stellar model
tracks of the subgiant branch to estimate precise ages and masses. 

We obtained Doppler measurements of the stars presented here using the
3\,m Shane and 0.6\,m Coude Auxiliary Telescope (CAT) at Lick
Observatory. Both telescopes feed the Hamilton 
spectrometer \citep{vogt87}, which has a resolution of $R \approx
50,000$ at $\lambda = 5500$~\AA. 
Doppler shifts are measured from each spectrum
using the iodine cell method described in detail by \citet{butler96}
and summarized as follows. A
temperature--controlled Pyrex cell containing 
gaseous iodine is placed at the entrance slit of the
spectrometer. The dense set of narrow molecular lines imprinted on
each stellar spectrum from 5000 to 6000~\AA\ provides a
robust wavelength scale for each observation, as well as information
about the shape of the spectrometer's instrumental response
\citep{marcy92b}.  

Traditionally, the Doppler shift of each stellar spectrum is measured
with respect to an observed, iodine--free template spectrum. 
These template observations require higher signal and resolution than 
normal radial velocity observations, which in turn necessitates longer
exposure 
times. Given our large target list and the small aperture of the CAT, 
obtaining a traditional template for each star would represent a
significant fraction of our allocated observing time, resulting in a
smaller than optimal target list. We therefore perform
a preliminary analysis of each star's observations using a
synthetic, ``morphed'' template spectrum following the method
described by \citet{johnson06a}. Stars showing conspicuous Doppler
variations are reanalyzed using a traditional template to
verify the signal and search for a full orbital solution. 

Internal uncertainties for each velocity measurement are estimated
from the standard deviation of the mean velocity measured from the
$\sim700$ segments analyzed in each echelle spectrum. A typical $V = 6$,
K0 star requires a 3600 second exposure on the CAT for a
signal-to-noise ratio (S/N) of 180 pix$^{-1}$, which produces an
internal precision of 
3--4~\ms. The same star observed on the Shane requires 300 seconds for
the same S/N owing to differences in the the collecting area and plate
scale at the spectrometer entrance slit. The radial velocities for the
\npllet\ stars presented here are listed in Table~\ref{tab1}
and Table~\ref{tab2}.

\section{Stellar Properties and Orbital Solutions} 
\label{stellar}

\begin{figure}[!t]
\epsscale{1}
\plotone{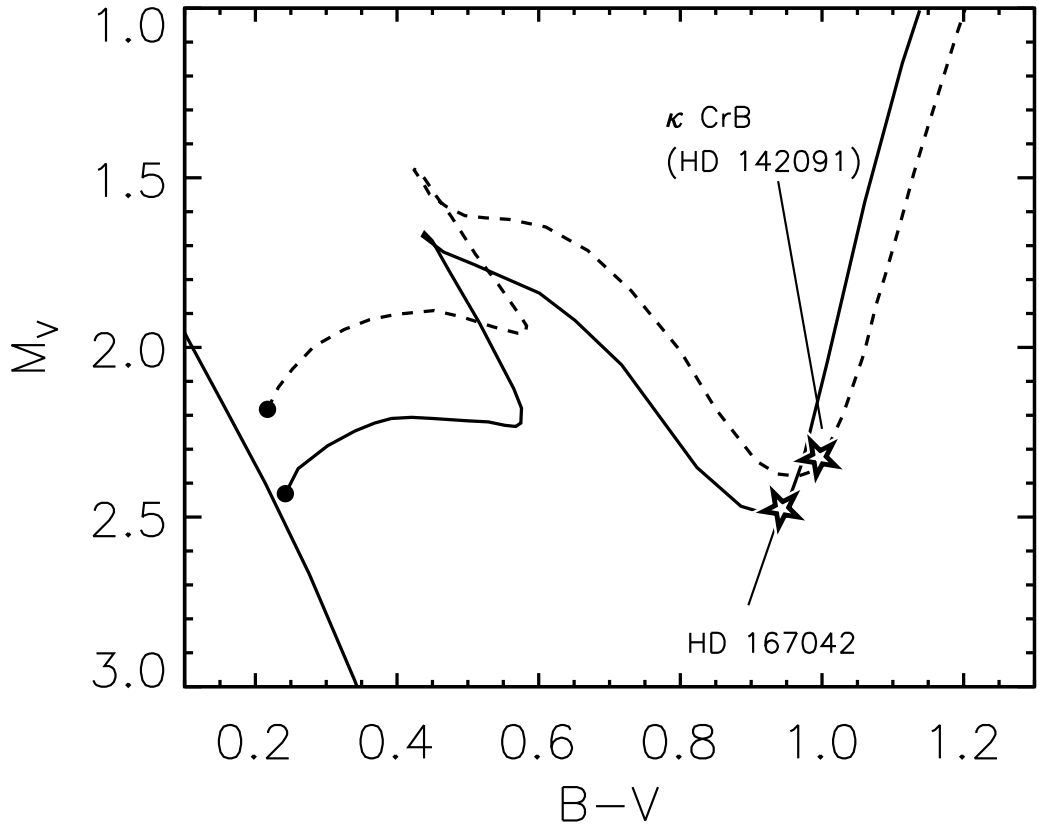}
\caption {\footnotesize{H--R diagram illustrating the properties of the
  \npllet\ subgiant planet host stars (pentagrams) compared to their
  main--sequence progenitors (filled circles). The connecting lines
  represent the \citet{girardi02} theoretical mass tracks
  interpolated for each star's metallicity. The thick, diagonal line
  is the zero--age main sequence assuming [Fe/H]=0.0.} \label{sg_hr}}    
\end{figure}

Our methods for determining the properties of our target stars are
described in \citet{johnson07} and summarized briefly as follows. We
use the LTE spectral synthesis code SME \citep{valenti05} to estimate
stellar effective temperatures, surface gravities,
metallicities, and projected rotational velocities by fitting a
synthetic spectrum to each star's iodine--free template
spectrum. We use the Stephan-Boltzmann Law to relate each star's 
radius and luminosity to its effective temperature,
parallax--based distance and bolometric correction. To estimate
stellar masses and ages, we interpolate each star's \emph{Hipparcos}
absolute magnitude, $B-V$ color and SME--derived metallicity onto the
stellar interior model grids computed by \citet{girardi02}. The
stellar properties and uncertainties are summarized in
Table~\ref{stellartable}. The position of the \npllet\ stars in the
H--R diagram is shown in Figure~\ref{sg_hr}, along with their
theoretical mass tracks and zero--age main sequence.

\citet{johnson07} estimate uncertainties of 7\% in stellar mass
and 1~Gyr for ages. Uncertainties in the SME--derived
parameters are given by \citet{valenti05}, however these estimates are
based primarily on a sample of main--sequence stars and may be
unrepresentative of the values obtained for our subgiants. We
therefore estimate uncertainties in the spectroscopic parameters by
exploring the degeneracy of the best--fit solution using different
input guesses to the SME code. We first solve for the best--fit
solution, and then use the resulting parameters as input for two additional
solutions, with $\pm 100$~K perturbations on $T_{eff}$. The
standard deviations of the parameters from the three SME trials are
then adopted as the $1\sigma$ parameter uncertainties. In cases when
our error estimates are less than those of \citet{valenti05}, we adopt the
latter values.

We search for the best fitting Keplerian orbital solution to each
radial velocity time series using a
Levenberg--Marquardt, least--squares minimization, and we estimate the 
uncertainties in the orbital parameters using a bootstrap 
Monte Carlo method. We first subtract the best--fit Keplerian from the
measured velocities. The residuals are then scrambled and added back
to the original measurements, and a new set of orbital parameters is
obtained. This process is repeated for 1000 trials and the standard
deviations of the parameters from all trials are adopted as
the formal 1$\sigma$ uncertainties. 

In the following subsections we describe the properties of the
\npllet\ new host stars and planet candidates discovered from our
sample of subgiants. 

\subsection{\starC}

$\kappa$ Coronae Borealis (\starC, HD\,142091, HR\,5901, HIP\,77655) is
listed in the SIMBAD database\footnote{http://cdsweb.u-strasbg.fr/}
with a spectral type K0IV and in the 
\emph{Hipparcos} catalog as a K0III-IV star with $V =
\vmagC$ and $B-V = \bvC$, and a parallax--based distance 
of \dC~pc \citep{hipp}. The star's apparent magnitude and distance
yield an absolute magnitude $M_V = \mvC$, placing the star
4.3~mag above the mean main sequence of stars in the Solar
neighborhood \citep{wright05}. The color and absolute magnitude of \starC\
suggest an evolved star on the subgiant branch near the base of the red
giant branch, in agreement with its published spectral classifications.

Our LTE spectral analysis suggests that
\starC\ is metal--rich, with 
[Fe/H]~$=\feC \pm \feeC$. Our spectral analysis also yields $T_{eff} =
\teffC \pm \teffeC$~K, 
\vsini~$=\vsiniC \pm 0.5$~\ks\ and surface gravity, $\log{g} = \loggC
\pm \loggeC$. We 
estimate the mass 
and age of \starC\ by interpolating the star's color, absolute 
magnitude and metallicity onto the \citet{girardi02} stellar interior
models. We find a stellar mass $M_* =$~\mstarC~\msun\ and an age of
\ageC~Gyr. We also estimate a luminosity $L_* =
\lstarC$~\lsun\ and radius $R_* = \rstarC$~\rsun, assuming a
bolometric correction -0.302. The properties of \starC\ are summarized
in Table~\ref{stellartable}.

To assess the photospheric stability of \starC, we searched the
\emph{Hipparcos} Epoch
Photometry
database\footnote{http://www.rssd.esa.int/index.php?project=HIPPARCOS} 
and found 135 observations. The measurements have a
mean uncertainty of 4.7~mmag and have an rms scatter of 4.7~mmag over
a time span of 3.1 years\footnote{After rejecting a single
  outlier}. Given its photometric stability, we can rule 
out significant pulsation modes that can contribute to intrinsic
radial velocity variability. Further, based on the lack of emission in
its CaII H\&K line core, we measure \rphk~$=\rhkC$ indicating
\starC\ is chromospherically inactive, similar to other evolved stars
\citep{wright04b}. 

\begin{figure}[!t]
\epsscale{1}
\plotone{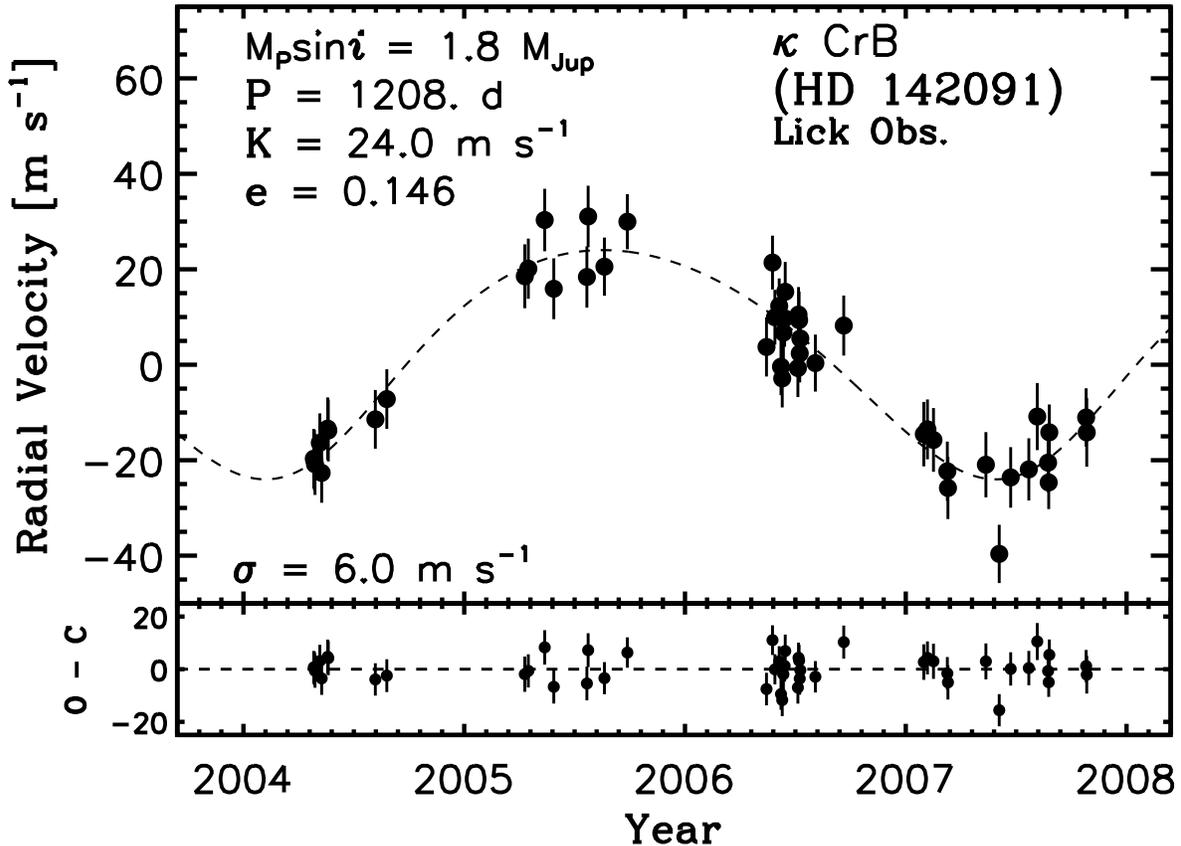}
\caption {Radial velocity time series for \starC\ (HD\,142091) measured at Lick
  Observatory. The dashed line shows the
  best--fit orbital solution, which has an orbital period of
  \pyearsC~years and \chisq~$ = \chisC$. \label{orbitC}}  
\end{figure}

Beginning in 2004 April, we  collected \nobsC\ Doppler
measurements at Lick Observatory, which are listed in
Table~\ref{tab1} together with their date of observation
and internal uncertainties. The velocities are also
shown in Figure~\ref{orbitC}, and the error bars represent the
quadrature sum of the internal measurement uncertainties and 5~\ms\ of
jitter, which is typical for the intrinsic Doppler variability of
subgiants similar to \starC\ \citep{johnson07}. 

Also shown in Figure~\ref{orbitC} is the best fitting Keplerian
orbital solution with a period $P = \pC \pm \peC$~days, eccentricity $e
= \eC \pm \eeC$ and velocity semiamplitude $K = \kC \pm
\keC$~\ms. The rms scatter of the data about the fit is \rmsC~\ms\ and
the reduced \chisq~$=\chisC$. Based on our stellar mass estimate of
$M_* = \mstarC$~\msun, the 
orbital solution gives a minimum planet mass \msini~$=\msiniC$~\mjup\
and semimajor axis $a = \arelC$~AU. The orbital parameters of
\starC b are summarized in Table~\ref{orbittable}. 

\subsection{\starA}

\starA\ (HR\,6817, HIP\,89047) is listed in the SIMBAD 
database with a spectral type K1\,III, suggesting the star is on the
giant branch. The
\emph{Hipparcos} catalog gives $V = 
\vmagA$, $B-V = \bvA$, and a parallax--based distance 
of \dA~pc \citep{hipp}. Given its distance and apparent magnitude, we
calculate an absolute magnitude $M_V = \mvA$, which 
places it 4.2~mag above the average main sequence of stars in the
Solar neighborhood \citep{wright05}. The position of \starA\ in the
H--R diagram indicates the star is better classified as a K1\,IV
subgiant near the upturn to the red giant branch, rather than a
luminosity class III giant. 

Based on our LTE spectral analysis \citep[SME;][]{valenti05}, we find
that \starA\ has Solar metal abundance with \feh~$=\feA
\pm \feeA$, and is 
slowly rotating with \vsini~$ = \vsiniA \pm 0.5$~\ks. Our spectroscopic
analysis also yields $T_{eff} = \teffA \pm \teffeA$~K and $\log{g} =
\loggA \pm \loggeA$.  
Interpolation of the star's color, absolute magnitude and metallicity
onto the \citet{girardi02} stellar model grids provides a 
stellar mass estimate of $M_* = \mstarA$~\msun, and an age of
\ageA~Gyr. Consistent with its post--main--sequence
evolutionary status, \starA\ is chromospherically inactive with
\rphk~$=$~\rhkA, as measured from its CaII~H\&K
emission relative to the stellar continuum. We also
estimate a radius $R_* = \rstarA \pm \rstareA$~\rsun\
and luminosity $L_* = \lstarA \pm \lstareA$~\lsun.

The \emph{Hipparcos} catalog lists 115
photometric measurements of \starA\ spanning 3.25 years. The star
is photometrically stable over this time baseline, with an rms scatter
11.2 mmag, which is slightly higher than the median measurement uncertainty
of 6 mmag. Since a periodogram analysis of the photometric
measurements shows no significant power at periods ranging from 2 to
1200 days, we expect that the contribution to the star's radial
velocity variability from star spots and radial pulsation should be
small. We follow \citet{johnson07} and adopt a jitter estimate of
5~\ms, based on the velocity scatter of other chromospherically quiet
subgiants with properties similar to \starA.  

\begin{figure}[!t]
\epsscale{1}
\plotone{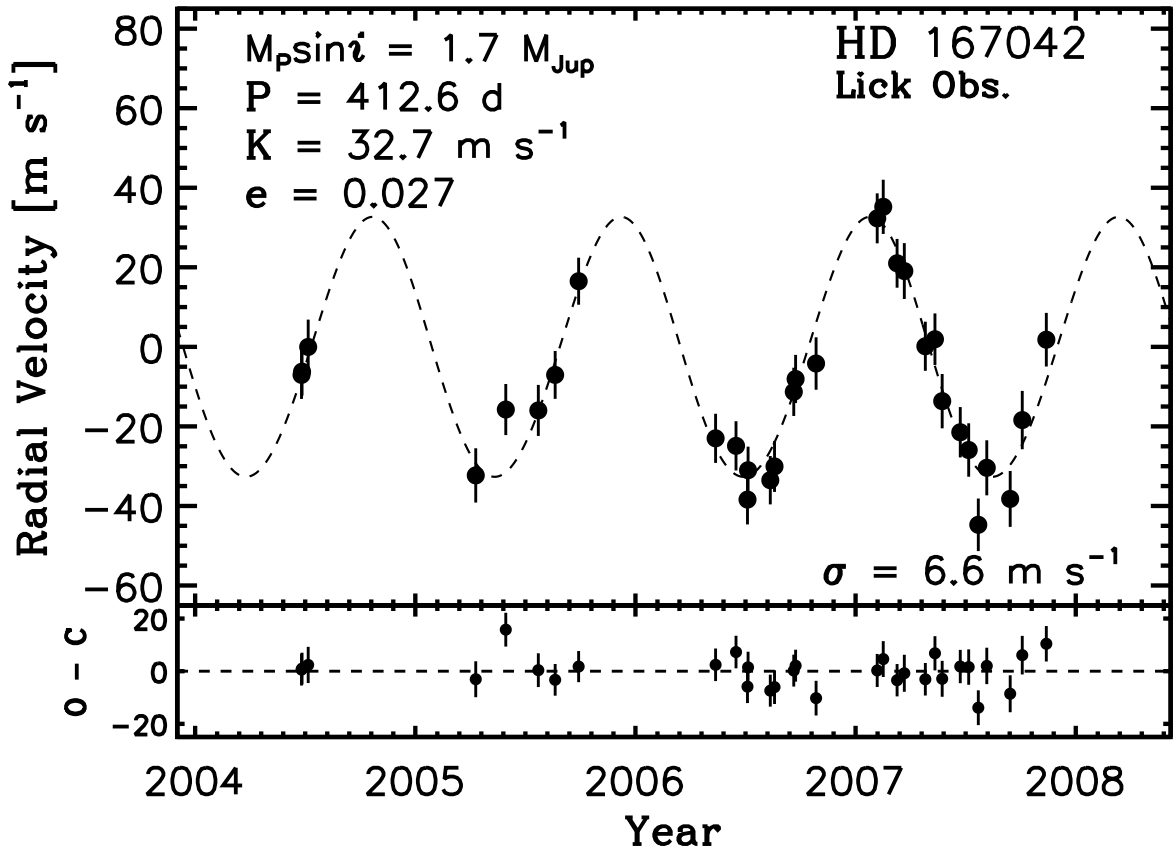}
\caption {Radial velocity time series for \starA\ measured at Lick
  Observatory. The dashed line shows the
  best--fit orbital solution, which has \chisq~$ = \chisA$. \label{orbitA}} 
\end{figure}

We began monitoring \starA\ in 2004 June and the
first 10 observations, analyzed with a synthetic
template, showed correlated variability with $\rm
rms=15$~\ms. We obtained a traditional, observed template
to confirm the variations with higher Doppler precision. The full set
of velocities is listed in Table~\ref{tab2} and
plotted in Figure \ref{orbitA}. The error bars in Figure~\ref{orbitA}
have been augmented by adding 5~\ms\ of jitter in quadrature to the
internal measurement uncertainties.

The best--fit Keplerian orbital solution is shown in
Figure~\ref{orbitA}. The solution has a
$\pA \pm \peA$~day period, a semiamplitude $K = \kA \pm \keA$~\ms, and
an eccentricity 
consistent with circular, $e = \eA \pm \eeA$. The residuals to the fit
have
$\rm rms = \rmsA$~\ms\ and reduced \chisq~$=$~\chisA, indicating that
the scatter is adequately modeled by the 
internal measurement uncertainties and estimated jitter. Assuming a
stellar mass 
$M_* = \mstarA$~\msun, the best--fit solution yields a semimajor axis $a
= \arelA$~AU, and minimum planet mass \msini~$=\msiniA$~\mjup. The
orbital parameters are summarized in Table~\ref{orbittable}.

\section{Summary and Discussion}
\label{summary}

We report the detection of \npllet\ Jovian planets orbiting
intermediate--mass subgiants. These detections come from our sample of
evolved stars that we are monitoring at Lick and Keck
Observatories. \starA\ and \starC\ have masses significantly larger
than solar, with $M_* = 
\mstarA$~\msun\ and  \mstarC~\msun, respectively. Examination of the
stars' theoretical mass tracks reveals that these current--day
subgiants began life on the main sequence as A--type
dwarfs (Figure~\ref{sg_hr}). 

The relatively low--amplitude Doppler variations induced by \starC b 
and \starA b would not have been detectable if the stars were
not in their current evolved state. Younger A-- and F--type
dwarfs have large rotational velocities (\vsini~$\gtrsim 50$~\ks) and
excessive pulsation--induced velocity jitter that can mask the reflex
velocity signal caused by a planet--mass companion
\citep{galland05}. On the other hand, older stars ascending the red giant
branch exhibit stochastic velocity variations in excess of 20~\ms
\citep{hekker06}, which would make the  planet orbiting \starC\
particularly difficult to detect since it induces a velocity amplitude
of only \kC~\ms. Subgiants, with their low 
rotation rates (\vsini~$ \lesssim 10$~\ks) and low jitter ($< 10$~\ms), 
occupy an observational ``sweet spot'' in the H--R diagram allowing
our Doppler survey to probe to relatively low planetary masses in a
wide range of orbital configurations.

The growing sample of planets around intermediate--mass stars is
beginning to reveal important relationships between stellar mass and
the properties of exoplanets. Of the 15
planets detected around evolved A--type stars ($M_* > 1.5$~\msun),
none have been found orbiting closer than $\sim 0.8$~AU, with the
majority orbiting at or beyond 1~AU \citep{johnson07}. \starC b and
\starA b are no 
exception to this trend, with semimajor axes of \arelC~AU and
\arelA~AU, respectively. As noted by \citet{johnson07},
this cannot be due to a decrease in detection sensitivity 
since, for a given planet and stellar mass, the amplitude of the
induced Doppler wobble scales as $a^{-1/2}$. Thus, the 
observed semimajor axis distribution of planets around A stars is
significantly different from that of planets around lower--mass 
stars. 

\citet{johnson07} considered the possibility that the lack of
close--in planets around K giants and clump giants 
may be attributable to engulfment by the expanding atmospheres of the
central stars. However, stars crossing the subgiant branch do
not undergo significant expansion, with radii smaller than $\sim
5$~\rsun\ even at the base of the red giant branch. The fact that none
of the 7 planets detected around massive subgiants ($M_* > 1.5$~\msun)
has $a < 0.9$~AU strongly suggests that the lack of
close--in planets around A stars is due to the effects of stellar mass
on planet formation and migration, rather than post--main--sequence
engulfment.

Stellar mass also plays an important role in the likelihood
that a star harbors a detectable giant planet. By measuring
the fraction of stars with planets in three widely spaced stellar mass
bins, \citet{johnson07b} found that the occurrence rate of 
planets with \msini~$>0.8$~\mjup\ and $a < 2.5$~AU is a rising
function of stellar mass. This analysis reveals
that A stars appear to be planet--enriched by a factor of 4.5  compared
to low--mass M dwarfs, with a measured giant planet occurrence rate of
9\% for $1.3 < M_*/M_\odot \leq 1.9$~\msun. The planet fraction for
higher--mass stars was based in 
part on three strong planet candidates from our survey of subgiants,
one of which is announced here (\starA b; \starC b
orbits beyond the 2.5 AU cut--off). This result
indicates that stellar mass is a strong tracer of
planeticity, which has important implications for the target selection
of future planet searches. Just as stellar metallicity is exploited in
the search for short--period planets around Sun--like
stars \citep{fischer05b, fischer05a, dasilva06},  
stellar mass should be an important consideration in the selection of
targets for future photometric, astrometric, and high--contrast direct
imaging surveys. 

The rising trend in planet occurrence toward higher stellar masses also
informs models of planet formation. Several theoretical studies of the
effects of stellar mass on 
planet formation have predicted that M stars should harbor
fewer Jovian planets than Sun--like dwarfs \citep{laughlin04,
ida05b}. More recently \citet{kennedy07} studied planet formation at
higher masses by accounting for the evolving disk mid--plane
temperature due to stellar irradiation and viscous evolution. They
predict that the fraction of stars with Jupiters should rise up to a
peak of $\approx 20$\% near $M_* = 3$~\msun. 

Testing this prediction requires searching for planets around stars
with masses greater than 2~\msun. Unfortunately, the mass range of
suitable subgiants is limited to $M* \lesssim 2.2$~\msun, primarily
due to our absolute magnitude criterion of $M_V > 2.0$ (see
\S~\ref{observations}). Doppler surveys of massive K giants with $M_*
> 2.5$~\msun\ provide perhaps the most favorable avenue for exploring
the occurrence rate of planets 
around more massive A stars \citep{frink01, sato05b,
  nied07}. In particular, \citet{lovis07} are
searching for planets around K giants in open clusters, which allows
them to take advantage of the uniform ages of the cluster members to
derive accurate stellar masses. The preliminary results from their
survey indicate that stars more massive than $\sim2$~\msun\ have an
enhanced abundance of super--Jupiters and brown dwarfs compared to
lower--mass stars.

To enlarge the statistical ensemble of planets detected around
intermediate--mass stars, we have expanded our planet search by adding
300 additional subgiants at Lick and Keck Observatories. If our
current 9\% detection rate holds, our expanded planet
search is expected to yield an additional 20-30 planets over the next
3 years. Together with the detections from Doppler surveys of K giants
and clump giants, the increased number of planets orbiting
intermediate--mass stars will add significantly to our understanding
of the effects of stellar mass on planet formation and planetary
system architecture.  

\acknowledgements 
We extend our gratitude to the many CAT observers who have helped 
with this project, including Chris McCarthy, Raj Sareen, Howard
Isaacson, Joshua Goldston, Bernie Walp, and Shannon Patel. We also
gratefully acknowledge the efforts and dedication of the Lick
Observatory staff, and the time assignment committee of the University
of California for their generous allocations of observing time. JAJ is
an NSF Astronomy and Astrophysics Postdoctoral Fellow and acknowledges
support from the NSF grant AST-0702821. We appreciate funding from 
NASA grant NNG05GK92G (to GWM). DAF is a Cottrell 
Science Scholar of Research Corporation and acknowledges support from
NASA Grant NNG05G164G that made this work possible. PKGW is supported
by an NSF Graduate Research Fellowship. This research has
made use of the SIMBAD database operated at CDS, Strasbourg France,
and the NASA ADS database. 


\clearpage
\begin{deluxetable}{lll}
\tablecaption{Radial Velocities for HD 142091\label{tab1}}
\tablewidth{0pt}
\tablehead{
\colhead{JD} &
\colhead{RV} &
\colhead{Uncertainty} \\
\colhead{-2440000} &
\colhead{(m~s$^{-1}$)} &
\colhead{(m~s$^{-1}$)} 
}
\startdata
13121.906 &  -24.17 &  3.84 \\
13122.948 &  -24.82 &  3.99 \\
13123.890 &  -24.02 &  4.06 \\
13131.838 &  -21.67 &  3.73 \\
13134.904 &  -26.75 &  3.79 \\
13144.858 &  -15.39 &  4.29 \\
13145.887 &  -15.79 &  4.11 \\
13223.735 &  -14.37 &  3.69 \\
13242.679 &   -9.87 &  3.75 \\
13470.970 &   18.04 &  4.43 \\
13476.824 &   16.88 &  3.83 \\
13503.906 &   30.83 &  4.39 \\
13518.921 &   13.77 &  3.90 \\
13573.710 &   12.21 &  4.01 \\
13575.763 &   27.63 &  4.27 \\
13602.694 &   17.93 &  3.45 \\
13640.657 &   28.65 &  2.89 \\
13870.807 &   -3.93 &  3.79 \\
13880.793 &   22.74 &  2.63 \\
13884.772 &   11.16 &  2.88 \\
13891.828 &   11.43 &  2.84 \\
13894.805 &   -6.89 &  3.14 \\
13896.825 &  -11.82 &  3.32 \\
13898.750 &    8.43 &  3.20 \\
13900.783 &    9.57 &  3.22 \\
13901.780 &   16.00 &  3.76 \\
13922.745 &   -2.59 &  3.44 \\
13923.811 &   10.59 &  2.91 \\
13924.765 &    6.76 &  3.32 \\
13925.805 &   -0.57 &  3.43 \\
13926.747 &    0.48 &  3.15 \\
13951.678 &   -4.63 &  3.34 \\
13998.623 &    5.82 &  3.94 \\
14131.082 &  -13.79 &  4.50 \\
14137.046 &  -16.25 &  3.84 \\
14147.025 &  -14.85 &  4.71 \\
14169.962 &  -26.75 &  3.66 \\
14171.015 &  -27.46 &  4.17 \\
14233.764 &  -21.38 &  4.84 \\
14255.856 &  -44.28 &  3.50 \\
14274.812 &  -25.60 &  3.89 \\
14304.755 &  -23.46 &  4.35 \\
14318.753 &  -13.02 &  4.92 \\
14336.656 &  -24.74 &  3.62 \\
14337.658 &  -26.46 &  2.43 \\
14338.661 &  -15.85 &  3.03
\enddata
\end{deluxetable}

\begin{deluxetable}{lll}
\tablecaption{Radial Velocities for HD 167042\label{tab2}}
\tablewidth{0pt}
\tablehead{
\colhead{JD} &
\colhead{RV} &
\colhead{Uncertainty} \\
\colhead{-2440000} &
\colhead{(m~s$^{-1}$)} &
\colhead{(m~s$^{-1}$)} 
}
\startdata
13181.937 &    5.64 &  3.40 \\
13182.835 &    6.45 &  3.36 \\
13192.971 &   12.66 &  4.71 \\
13471.006 &  -19.63 &  4.64 \\
13520.854 &   -3.08 &  4.03 \\
13574.796 &   -3.30 &  4.02 \\
13602.721 &    5.63 &  3.36 \\
13641.710 &   29.19 &  3.16 \\
13868.944 &  -10.32 &  3.64 \\
13902.825 &  -12.20 &  3.64 \\
13921.848 &  -25.72 &  3.73 \\
13922.772 &  -18.33 &  3.18 \\
13959.673 &  -20.84 &  3.46 \\
13966.672 &  -17.38 &  4.00 \\
13998.646 &    1.37 &  3.44 \\
14001.669 &    4.61 &  3.42 \\
14035.620 &    8.49 &  4.29 \\
14137.078 &   44.99 &  3.81 \\
14147.063 &   47.89 &  4.63 \\
14170.039 &   33.69 &  3.56 \\
14182.036 &   31.74 &  4.94 \\
14216.930 &   12.83 &  3.63 \\
14232.868 &   14.60 &  4.14 \\
14244.873 &   -0.97 &  4.65 \\
14274.859 &   -8.79 &  3.88 \\
14288.829 &  -13.25 &  4.49 \\
14304.764 &  -32.06 &  4.32 \\
14318.828 &  -17.73 &  4.78 \\
14357.668 &  -25.54 &  4.93 \\
14377.725 &   -5.72 &  5.36 \\
14417.653 &   14.48 &  4.52
\enddata
\end{deluxetable}

\begin{deluxetable}{lll}
\tablecaption{Stellar Parameters\label{stellartable}}
\tablewidth{0pt}
\tablehead{
  \colhead{Parameter} & 
  \colhead{\starC\tablenotemark{a}}    &
  \colhead{\starA}    
}
\startdata
V               & \vmagC        & \vmagA              \\
$M_V$           & \mvC          & \mvA                \\
B-V             & \bvC          & \bvA                \\
Distance (pc)   & \dC           & \dA                 \\
${\rm [Fe/H]}$  & \feC~(\feeC)   & \feA~(\feeA)       \\
$T_{eff}$~(K)   & \teffC~(\teffeC) & \teffA~(\teffeA) \\
\vsini~(\ks)    & \vsiniC~(0.5) & \vsiniA~(0.5)       \\
$\log{g}$       & \loggC~(\loggeC) & \loggA~(\loggeA) \\
$M_{*}$~(\msun) & \mstarC~(0.11)& \mstarA~(0.13)      \\
$R_{*}$~(\rsun) & \rstarC~(0.08)& \rstarA~(0.07)      \\
$L_{*}$~(\lsun) & \lstarC~(0.04)& \lstarA~(0.05)      \\
Age~(Gyr)       & \ageC~(1.0)   & \ageA~(1.0)         \\
$S_{HK}$        & \shkC         & \shkA               \\
$\log R'_{HK}$  & \rhkC         & \rhkA               \\
\enddata
\tablenotetext{a}{HD\,142091}
\end{deluxetable}

\clearpage
\begin{deluxetable}{llll}
\tablecaption{Orbital Parameters\label{orbittable}}
\tablewidth{0pt}
\tablehead{\colhead{Parameter} &
  \colhead{\starC b\tablenotemark{a}}    &
  \colhead{\starA b}    
}
\startdata
P (d)         & \pC~(\peC)   & \pA~(\peA)     \\
T$_p$\tablenotemark{b}~(JD)    & \tpC~(\tpeC) & \tpA~(\tpeA) \\
e             & \eC~(\eeC)   & \eA~(\eeA)     \\
K~(\ms)   & \kC~(\keC)   & \kA~(\keA)         \\
$\omega$~(deg)& \omC~(\omeC) & \omA~(\omeA) ) \\
\msini~(\mjup)& \msiniC      & \msiniA        \\
$a$~(AU)      & \arelC       & \arelA         \\
Fit RMS~(\ms) & \rmsC        & \rmsA          \\
Jitter~(\ms)  & 5.0          & 5.0            \\
\chisq        & \chisC       & \chisA         \\
N$_{\rm obs}$ & \nobsC       & \nobsA         \\
\enddata
\tablenotetext{a}{HD\,142091\,b}
\tablenotetext{b}{Time of periastron passage.}
\end{deluxetable}

\end{document}